\documentclass[twocolumn,prl,aps,superscriptaddress,showpacs]{revtex4}

\usepackage{dcolumn}
\usepackage{amsmath}
\usepackage{epsfig}

\newlength{\figwidth}
\newlength{\figwidthb}
\begin{document}

\title{Non-complicated $\rm EuTiO_3$ Structure}
\author{David S. Ellis}
\email{david_ellis@spring8.or.jp}
\affiliation{Materials Dynamics Laboratory, RIKEN SPring-8 Center, RIKEN, 1-1-1 Kouto, Sayo, Hyogo, 679-5148, Japan}
\author{Hiroshi Uchiyama}
\affiliation{Materials Dynamics Laboratory, RIKEN SPring-8 Center, RIKEN, 1-1-1 Kouto, Sayo, Hyogo, 679-5148, Japan}
\affiliation{Research and Utilization Division, SPring-8/JASRI, Sayo, Hyogo, 679-5198, Japan}
\author{Satoshi Tsutsui}
\affiliation{Research and Utilization Division, SPring-8/JASRI, Sayo, Hyogo, 679-5198, Japan}
\author{Kunihisa Sugimoto}
\affiliation{Research and Utilization Division, SPring-8/JASRI, Sayo, Hyogo, 679-5198, Japan}
\author{Kenichi Kato}
\affiliation{Structural Materials Science Laboratory, RIKEN SPring-8 Center, 1-1-1 Kouto, Sayo, Hyogo, 679-5148, Japan}
\author{Alfred Q. R. Baron}
\email{baron@spring8.or.jp}
\affiliation{Materials Dynamics Laboratory, RIKEN SPring-8 Center, RIKEN, 1-1-1 Kouto, Sayo, Hyogo, 679-5148, Japan}
\affiliation{Research and Utilization Division, SPring-8/JASRI, Sayo, Hyogo, 679-5198, Japan}

\date{\today}

\begin{abstract}
\end{abstract}


\maketitle

In a recent Letter on $\rm EuTiO_3$, incommensurate reflections were observed as satellites to the \textbf{q}=(0.5 0.5 0.5) \emph{R}-point, in some cases much stronger than the \emph{R}-point reflection itself \cite{Kim12}.  This modulation was attributed to competing antiferrodistortive and antiferroelectric phases.  While this observation allows for interesting modeling to be done, we wish to point out that such incommensurabilities should not be taken as the general case for this material.   They have only been observed in samples from one source, and there are other published studies where no such incommensurabilites are observed.  We discuss this in detail and present supporting data below.\\

\vspace{-1 mm}

There have been several recent diffraction studies \cite{Allieta12,Kohler12,Goian12,Ellis12,Ellis12b}. The powder measurements of \cite{Allieta12,Kohler12} show a transition into a structure consistent with a low temperature I4/\emph{mcm} phase, but we cannot draw conclusions about the presence of satellites from the presented data.  Goian et al. \cite{Goian12} measured electron and x-ray diffraction of ceramic and single crystal $\rm EuTiO_3$.  In the reconstructed electron diffraction patterns from ceramic samples fabricated as described in \cite{Kachlik12}, superlattice \emph{R}-point reflections were observed \emph{without} the incommensurate satellites. The single crystal samples, which were obtained from the same source as used in \cite{Kim12}, showed four-fold satellite peaks in electron diffraction patterns.  The satellites in \cite{Goian12} were (0.12 $\pm$ 0.02) offset from the \emph{R} point, compared to 0.07 in \cite{Kim12}, in units of reciprocal lattice.  Meanwhile, our $\rm EuTiO_3$ samples \cite{Ellis12,Ellis12b} showed no evidence at all of satellites, neither in single crystal nor in powder diffraction.  As a specific check, figure 1 shows one of our early single crystal diffraction measurements at \emph{T}=250 K.   In the figure, the \emph{R}-point peak is clearly present at \textbf{Q}=(0.5 2.5 -1.5), but there is no trace of the satellites described in \cite{Kim12}.\\

\vspace{-2 mm}

Considering samples from three independent sources, two of the three do not show the incommensurate peaks seen in \cite{Kim12}.  And within the group of samples where the incommensurabilities exist, there are discrepencies in the incommensurate wave-vector.  Therefore, we suggest that a simple, un-modulated, structure is the norm for $\rm EuTiO_3$.

  \begin{figure}
  \centering
  \epsfig{file=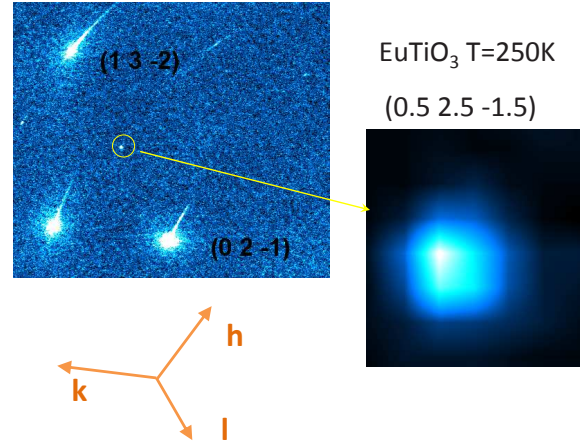,height=2.3in,keepaspectratio}

  \caption{(Color Online)   Single crystal x-ray diffraction pattern of a small piece ($\sim$ 5 $\mu$m) of single crystal $\rm EuTiO_3$ at \emph{T}=250 K, measured at BL02B1 of SPring-8 synchrotron, showing the \textbf{Q}=(0.5 2.5 -1.5) \emph{R}-point midway between the (1 3 -2) and (0 2 -1) peaks.  The approximate \textbf{h},\textbf{k}, and \textbf{l} projections on the image plane are indicated. } \label{fig:diffract2D}
  \end{figure}

\vspace{-5 mm}


\end{document}